\documentclass [prl, twocolumn, aps, superscriptaddress
] {revtex4}
\usepackage{graphicx}
\usepackage{amsmath}
\usepackage{amssymb}

\newcommand{\vect}[1]{{\mathbf #1}}

\begin{document}

\title{Magnetocapacitance in non-magnetic inhomogeneous media}

\author{Meera M. Parish}
\email{mparish@princeton.edu} %
\affiliation{Department of Physics, Princeton University, Princeton,
NJ 08544} %
\affiliation{Princeton Center for Theoretical Science, Princeton
University, Princeton, NJ 08544}

\author{Peter B. Littlewood}
\affiliation{Cavendish Laboratory, Madingley Road, Cambridge CB3
0HE, United Kingdom}

\date{\today}

\begin{abstract}
The dielectric response in a magnetic field is routinely used to
probe the existence of coupled magnetic and elastic order in the
multiferroics. However, here we demonstrate that magnetism is not
necessary to produce a magnetocapacitance when the material is
inhomogeneous. By considering a two-dimensional, two-component
composite medium, we find a characteristic dielectric resonance that
depends on magnetic field. We propose this as a possible signature
of inhomogeneities and we argue that this behavior has already been
observed in nanoporous silicon and some manganites.
\end{abstract}



\maketitle

The behavior of a material's dielectric constant under an applied,
alternating electric field provides a powerful probe of its
microscopic properties. For example, dielectric relaxation
(characterized by a delay in the electrical polarization with
respect to the changing electric field) can signify the presence of
non-interacting, freely-rotating dipoles~\cite{jonscher1983}. More
recently, the dielectric response as a function of an applied static
magnetic field has been used to determine whether or not a material
is a
\emph{multiferroic}~\cite{hemberger2003,kimura2003,hur2004,singh2005}. 
In multiferroic systems, magnetic order and electrical polarization
are coupled, thus giving rise to a magnetocapacitance. This
magneto-electric coupling makes multiferroics a topic of interest
owing to its potential application~\cite{eerenstein2006}.

It is well known that Maxwell-Wagner extrinsic effects, such as
inhomogeneities and contact effects, can enhance the dielectric
constant and yield dielectric relaxation in the absence of intrinsic
dipolar relaxation~\cite{maxwell91,wagner1913,lunkenheimer2002}.
However, it was only recently appreciated that the Maxwell-Wagner
effect can also yield a magnetocapacitance \emph{without}
multiferroicity and its associated magnetoelectric coupling,
provided the material exhibits an intrinsic
magnetoresistance~\cite{catalan2006}. Such effects clearly
demonstrate that a magnetocapacitance is not sufficient to establish
multiferroicity. On the other hand, having a magnetocapacitance
without magnetoelectric coupling may be more practical for
technological applications.

In this letter, we show that in an inhomogeneous medium one does not
require even an intrinsic magnetoresistance to produce a
magnetic-field-dependent dielectric constant. Thus, there is the
possibility of a sizable magnetocapacitance without any magnetic
order.

As far as we are aware, our work is the first to address the
dielectric response of composite media in the presence of a magnetic
field, in a regime relevant to semiconductors. While the dc
magnetotransport in classical, disordered media has been studied
extensively~\cite{I92,herring,dreizin,stroud,bala,me,parish2005}, 
there are very few works on the ac dielectric response. Moreover,
the latter are restricted to metal-insulator composites which are
either in the absence of a magnetic field~\cite{murtanto2006} or
they have an electric-field frequency $\omega$ that is close to the
plasma resonance frequency~\cite{bergman1998}. Such a regime is only
applicable to highly metallic materials rather then the
semiconducting compounds we are interested in here.

We find that the dielectric function of a composite medium exhibits
 unexpected behavior in a magnetic field and, as such, could be
used as a probe for inhomogeneities. Specifically, we find a strong
dielectric resonance as a function of frequency, where the position
of the dissipation peak sensitively depends on magnetic field.
Furthermore, the real part of the dielectric response even switches
sign in the vicinity of the resonance at finite field. Using these
results, we argue that inhomogeneities are a probable cause of the
magnetic-field-dependent `dielectric relaxation' observed in
nanoporous silicon~\cite{vasic2007} and in some manganite
compounds~\cite{rivas2006,rairigh2007}.

To focus our investigation, we restrict ourselves to a
two-component, two-dimensional (2D) composite medium. This should be
sufficient to describe the salient features of a disordered
material, but simple enough for us to derive analytical expressions.
We assume that the length scale of the inhomogeneities is much
larger than all the microscopic length scales of the system, e.g.\
the mean free path.
Thus, the local current density $\vect{j}(\omega,\vect{r})$ is
related to the local electric field $\vect{E}(\omega,\vect{r})$ via
Ohm's law: $\vect{j}(\omega,\vect{r}) =
\hat\sigma(\omega,\vect{r})\vect{E}(\omega,\vect{r}) \equiv i\omega
\hat\varepsilon(\omega,\vect{r})\vect{E}(\omega,\vect{r})$, where
$\hat\sigma(\omega,\vect{r})$ is the local conductivity tensor and
$\hat\varepsilon(\omega,\vect{r})$ is the local dielectric function.
Furthermore, we take the regime where (i) $1/\omega$ is much larger
than the scattering time, and (ii) the intrinsic conductivity and
dielectric constant of each component are frequency independent.
Then, the conductivity tensor in a transverse magnetic field
$\vect{H} = H\hat z$ is given by
%
\begin{align}\label{eq:sigma}
\hat\sigma(\omega) =
 \frac{\sigma}{1 + \beta^2}
 \left (
\begin{array}{ccc}
1 & \beta &  \\
-\beta & 1 & 
\end{array}
\right ) + i\omega\varepsilon \mathbf{1}
\end{align}
%
%
Here, $\sigma$ is the dc scalar conductivity, $\varepsilon$ is the
bare dielectric constant and $\beta = \mu H$, where $\mu$ is the
carrier mobility. In general, the quantities $\sigma$, $\varepsilon$
and $\mu$ will be random functions of space, but here we will only
allow them to take on two different values.

\begin{figure}
\centering
\includegraphics[width = 0.25\textwidth]{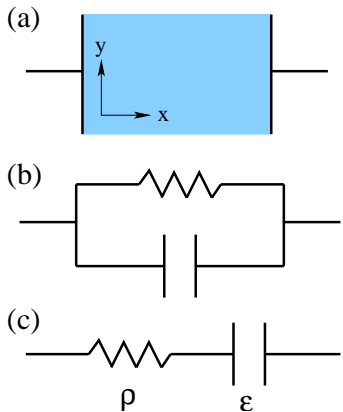}
\caption{\label{fig:circuits} Simple circuits of resistors $\hat\rho
\equiv \hat\sigma^{-1}$ and capacitors $\varepsilon$ that illustrate
the basics of dielectric response. Diagram (a) depicts the measuring
set-up, where a rectangular sample is confined between metallic
plates and subjected to an applied electric field $E_x \hat x$, (b)
represents a simple homogeneous medium, and (c) represents the
simplest metal-insulator composite that exhibits the Maxwell-Wagner
effect.}
\end{figure}

By taking volume averages over the composite medium, we can define
an effective conductivity such that $\langle\vect{j}(\omega)\rangle
= \hat\sigma_e(\omega) \langle\vect{E}(\omega)\rangle$.
However, for the typical measuring set-up depicted in
Fig.~\ref{fig:circuits}(a), we clearly have the boundary condition
$j_y = 0$, and so it is more natural to use the effective
resistivity $\hat\rho_e(\omega) \equiv \hat\sigma_e(\omega)^{-1}$.
Thus, the components of the dielectric function that are actually
probed in experiment are given by $\varepsilon_{xx}(\omega) =
(i\omega\rho_{e,xx}(\omega))^{-1}$ and $\varepsilon_{xy}(\omega) =
(i\omega\rho_{e,xy}(\omega))^{-1}$.
The longitudinal response $\varepsilon_{xx}(\omega)$ is simply the
capacitance of the configuration in Fig.~\ref{fig:circuits}(a),
while the transverse response $\varepsilon_{xy}(\omega)$ can be
extracted from a measurement of the transverse electric field
$E_y(\omega) =
(\varepsilon_{xx}(\omega)/\varepsilon_{xy}(\omega))E_x(\omega)$.
Note that in the limit $\beta \rightarrow 0$, we must have
$E_y(\omega) \rightarrow 0$ and thus $\varepsilon_{xy}(\omega)
\rightarrow \infty$.

First, it is instructive to consider a homogeneous medium described
by the conductivity tensor in Eq.~\eqref{eq:sigma}. One can
represent this as a capacitor in parallel with a resistor (see
Fig.~\ref{fig:circuits}(b)), where both circuit elements are
frequency-independent. For this simple case, we find the real part
of the longitudinal dielectric response to be
\begin{equation}
 \Re[\varepsilon_{xx}(\omega)] = \frac{\varepsilon(1 - \beta^2 + (\omega\tau)^2(1+\beta^2)^2)}
 {1 + (\omega\tau)^2(1+\beta^2)^2}
\end{equation}
where $\tau = \varepsilon/\sigma \equiv \rho\varepsilon$. Thus, we
see that $\Re[\varepsilon_{xx}(\omega)]$ can be \emph{negative} at
sufficiently low frequencies when $\beta >1$.
Usually, in the absence of a magnetic field,
$\Re[\varepsilon_{xx}(\omega)]$ only goes negative at a dielectric
resonance, and this signifies the presence of an inductive element
in the system.
However, for our simple resistor-capacitor circuit in a strong
magnetic field, the boundary condition $j_y = 0$ causes the Hall
component of the resistive element to be mixed into the longitudinal
dielectric response, resulting in a negative real part.

Now lets switch off the magnetic field and consider the simplest
realization of an inhomogeneous medium: a resistor and capacitor in
series (Fig.~\ref{fig:circuits}(c)). Here, we have dielectric
function:
\begin{align}
\varepsilon_{xx}(\omega) & = \frac{\varepsilon (1 - i\omega\tau)}{1
+ (\omega\tau)^2}
\end{align}
For low frequencies $\omega\tau < 1$, the capacitor dominates the
response, while at high frequencies $\omega\tau >1$, the voltage
drop falls primarily over the resistor instead since there is
insufficient time for the capacitor to build up charge.
Thus, at the characteristic frequency $\omega\tau = 1$, we have a
rapid change in $\Re[\varepsilon_{xx}(\omega)]$ and an associated
peak in $\Im[\varepsilon_{xx}(\omega)]$.
This is a basic illustration of the Maxwell-Wagner effect, where the
inhomogeneity mixes the real and imaginary modes of the response to
create an apparent dielectric relaxation.

From these simple circuits, we see that geometric effects and
macroscopic inhomogeneities can mix the different `modes'
(longitudinal, transverse, real and imaginary) of the system, and
thus generate unexpectedly rich behavior, such as the dielectric
resonance we will describe below. Inhomogeneities play a similar
role in the dc magnetotransport of heavily disordered media, since
they can mix the Hall resistivity into the longitudinal component to
produce a linear magnetoresistance~\cite{me}.

We now address the full problem of 2D isotropic inhomogeneous media.
We suppose that the material is composed of two phases
$\hat\sigma_1(\omega)$ and $\hat\sigma_2(\omega)$ in proportions
$p_1=p$ and $p_2=1-p$, respectively. To model the dielectric
response of a strongly inhomogeneous medium, we take the extreme
case $\sigma_1 = 0$, $\varepsilon_1 = \varepsilon$ and $\sigma_2 =
\sigma$, $\varepsilon_2 = 0$, i.e.\ we consider purely capacitive
and purely resistive regions.
When the proportions of each phase are equal ($p=1/2$), we can use a
symmetry transformation for the current density and electric
field~\cite{D71,B78,DR94} to derive the \emph{exact} result for the
components of the effective dielectric function:
\begin{align}\label{eq:effdiel}
\varepsilon_{xx}(\omega) & = \varepsilon \frac{(1+i\omega\tau)}
{\sqrt{i\omega\tau}\sqrt{(1+i\omega\tau)^2-(\omega\tau\beta)^2}}
 \\ 
\varepsilon_{xy}(\omega) & =
\frac{\varepsilon}{\beta}\left(1-\frac{i}{\omega\tau}\right)
\end{align}
We immediately see that the Hall component
$\varepsilon_{xy}(\omega)$ is equivalent to the longitudinal
dielectric response of a homogeneous material
(Fig.~\ref{fig:circuits}(b)) in the absence of a magnetic field,
with bare dielectric constant $\varepsilon/\beta$ and resistivity
$\beta\rho$ (which is just the Hall resistivity for dc
magnetotransport). As we will see below, this is unique to the case
where the proportions are equal.

\begin{figure}
\centering
\includegraphics[width = 0.48\textwidth]{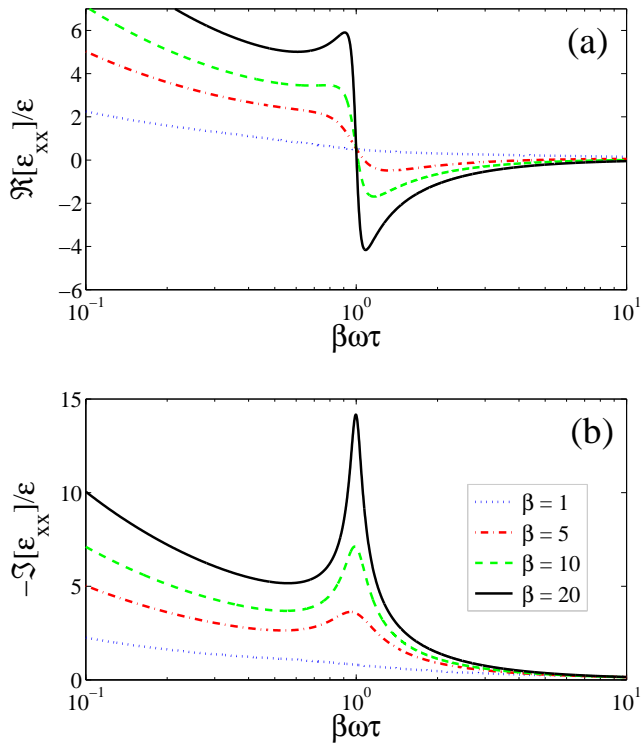}
\caption{\label{fig:dielH05}Dielectric response versus frequency
over a range of magnetic fields $\beta$ for a 2D two-component
medium with equal proportions ($p=1/2$). When $\beta > 1$, there is
a resonance at normalized frequency $\beta\omega\tau = 1$, where the
real part (a) varies rapidly, eventually changing sign, while the
imaginary part (b) exhibits a peak.}
\end{figure}

The longitudinal dielectric response $\varepsilon_{xx}(\omega)$, on
the other hand, involves a non-trivial mixture of dielectric and
resistive components in both the real and imaginary parts. The
behavior is plotted in Fig.~\ref{fig:dielH05} and represents the key
result of this paper.
For large magnetic fields $\beta > 1$, or, equivalently, for small
frequencies $\omega\tau <1$, a dielectric resonance occurs at
$\beta\omega\tau = 1$, where $\Im[\varepsilon_{xx}(\omega)]$ has a
pronounced peak and $\Re[\varepsilon_{xx}(\omega)]$ varies rapidly.
In particular, we see that the resonant frequency is determined by
the Hall resistivity $\beta\rho$, instead of $\rho$ like in the
usual Maxwell-Wagner effect. Thus, the position of the peak in
frequency space is inversely proportional to $\beta$.
%
By substituting $\beta\omega\tau = 1$ into Eq.~\eqref{eq:effdiel},
we see that the peak height
$\Im[\varepsilon^p_{xx}] \simeq 
\varepsilon\beta/\sqrt{2}$. However, the actual dissipation at the
peak remains constant because here we always have $\Re[\rho_{e,xx}]
\simeq \sqrt{2} \rho$.
In addition, the real part $\Re[\varepsilon_{xx}(\omega)]$ becomes
\textit{negative} for $\beta\omega\tau \gtrsim 1$ and $\beta > 1$,
but it exhibits ordinary Maxwell-Wagner dielectric relaxation at
zero magnetic field.
Surprisingly, there is no corresponding peak in
$\Im[\varepsilon_{xx}(\omega)]$ when $\beta = 0$, and this holds for
arbitrary insulator fraction $p$ according to results obtained from
the effective medium approximation and numerical resistor network
studies~\cite{murtanto2006}. This is particularly unexpected given
that we obtain a dielectric relaxation peak from the simplest
metal-insulator composite in Fig.~\ref{fig:circuits}(c).
However, we speculate that one may recover the peak when the
disorder is anisotropic since Fig.~\ref{fig:circuits}(c) is
obviously highly anisotropic.

\begin{figure}
\centering
\includegraphics[width = 0.42\textwidth]{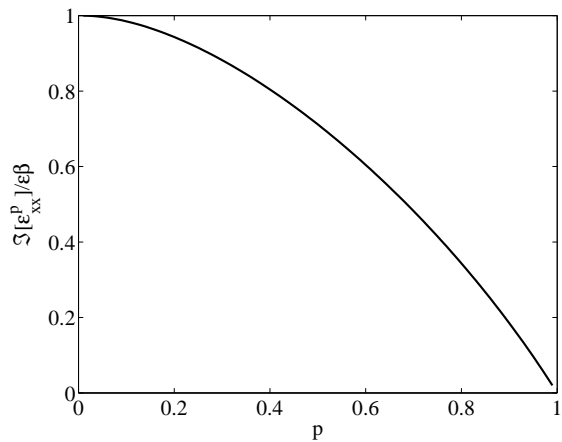}
\caption{\label{fig:peak} Behavior of the resonance peak
$\Im[\varepsilon^p_{xx}]$ at $\beta\omega\tau = 1$ as a function of
the fraction $p$ of the insulating phase. This holds for all $\beta
> 1$.}
\end{figure}

To determine whether or not this dielectric resonance is a generic
feature of 2D conductor-insulator composites, we use the effective
medium approximation~\cite{S75,GS05} to examine the case where
$p\neq 1/2$. This amounts to numerically solving the coupled
equations:
\begin{align}
\sum_i p_i (\hat\sigma_i - \hat\sigma_e)
\left(\mathbf{1}-\frac{\hat\sigma_i -
\hat\sigma_e}{2\sigma_{e,xx}}\right)^{-1} & = 0
\end{align}
When $\beta > 1$, we find that there is always a resonance at
$\beta\omega\tau = 1$ for arbitrary $p$, but that the height of the
dissipation peak $\Im[\varepsilon^p_{xx}(\omega)]$ decreases with
increasing $p$ (decreasing conductor fraction) as shown in
Fig.~\ref{fig:peak}.
Also, the behavior of $\Re[\varepsilon_{xx}(\omega)]$ is
qualitatively unchanged at the resonance, apart from the fact that
the point at which it switches sign occurs at a higher field for
increasing $p$. This makes sense given that it is the Hall component
of the conducting region that is responsible for the negative part
of $\Re[\varepsilon_{xx}(\omega)]$.

\begin{figure}
\centering
\includegraphics[width = 0.46\textwidth]{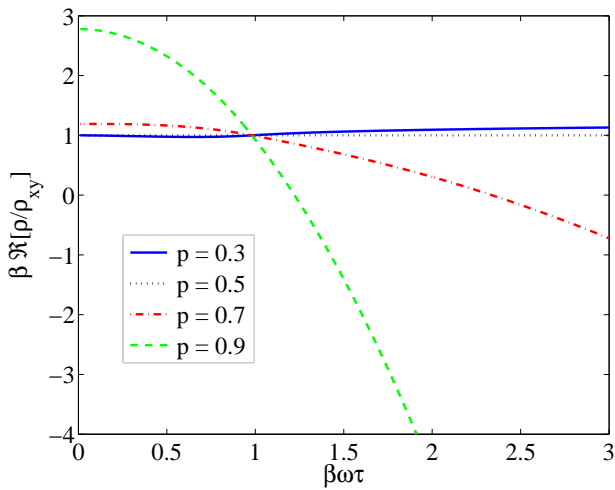}
\caption{\label{fig:sigxyH5} Hall component $\Re[\rho/\rho_{e,xy}]
\equiv -\omega\tau \Im[\varepsilon_{xy}]/\varepsilon$ at $\beta = 5$
as a function of frequency for different fractions $p$ of the
insulating phase. Similar curves are obtained for other $\beta > 1$.
At $\beta\omega\tau = 1$, $\Im[\varepsilon_{xy}]/\varepsilon = 1$
for all $p$ and $\beta$.}
\end{figure}

The largest variation with respect to $p$ occurs in the Hall
component $\Im[\varepsilon_{xy}(\omega)]$. As shown in
Fig.~\ref{fig:sigxyH5}, while $\Re[\rho/\rho_{e,xy}]$ is relatively
insensitive to frequency for $p \leq 1/2$, it goes negative beyond a
critical value of $\beta\omega\tau > 1$ when $p > 1/2$. Indeed, in
the limit $p \rightarrow 1$, we have $\Re[\rho/\rho_{e,xy}]
\rightarrow \infty$ for $\beta\omega\tau < 1$ and
$\Re[\rho/\rho_{e,xy}] \rightarrow -\infty$ for $\beta\omega\tau >
1$, with $\Re[\rho/\rho_{e,xy}] = 1/\beta$ always at
$\beta\omega\tau = 1$.
Thus, we see there is some symmetry to the `mode-mixing': the
dielectric region reverses the sign of the Hall component at high
frequencies in the same way that the Hall resistivity forces the
dielectric constant to become negative at low frequencies.
In principle, one could exploit this property to estimate the
insulator fraction $p$ of a composite medium.

Finally, we emphasize that the existence of the dielectric resonance
is not conditional on there being a magnetoresistance at $\omega =
0$. One clearly sees this from the fact that there is always a
resonance at $\beta\omega\tau = 1$ when $\beta > 1$, even though the
magnetoresistance is zero for $p<1/2$ and undefined for $p\geq 1/2$.
One can obtain a non-zero magnetoresistance when the dielectric
region has a non-zero conductivity $\sigma_1 \neq \sigma_2$, but
this saturates with magnetic field unless the proportions are equal
($p = 1/2$)~\cite{GS05,magier2006}.

The magnetic-field-dependent dielectric resonance described above
has already been observed experimentally in nanoporous
silicon~\cite{vasic2007}, a material which clearly resembles a
conductor-insulator composite. Specifically, the dielectric response
was measured as a function of temperature $T = 25-40$K at fixed
frequency $\omega =$ 100kHz and magnetic field $H = 0-32$T,
but this is equivalent to varying $\omega\tau$ if we assume that
$\varepsilon$ is temperature-independent and that the electrical
transport in the semiconductor is activated: $\rho = \rho_0
e^{\Delta/k_BT}$, where $\Delta$ is the band gap.
Then the temperature at which the resonance occurs is $k_B T =
\Delta/\log(1/\beta \omega \varepsilon \rho_0)$, so that we expect
the dielectric resonance to shift to higher $T$ with increasing $H$,
as indeed was observed. Assuming that $H=$ 1T corresponds to $\beta
\simeq 1$, we can fit the data with $\Delta \simeq 30$meV and
$\omega_0 \equiv 1/\rho_0\varepsilon \simeq 8\times 10^{10}$Hz. Our
results should be contrasted with an alternative
proposal~\cite{Brooks2008} which interprets the data with a model of
Debye relaxation, requiring that charge carriers are pinned to a
distribution of sites, but which does not explicitly introduce
inhomogeneous fields.

In principle, one can exploit the dielectric resonance to construct
a magnetic sensor that is sensitive to fields in the neighborhood of
$\beta \sim 1/\omega\tau$.
However, the challenge is to make it operational within the ideal
frequency range $\omega = 1$kHz$- 1$MHz at room temperature and for
$H < 1$T, since this requires a semiconductor that has both a large
band gap $\Delta > 0.5$eV and a high mobility $\mu > 1$T$^{-1}$.

A dielectric resonance has also been observed in the manganite
La$_{2/3}$Ca$_{1/3}$MnO$_3$ at \emph{zero} magnetic field, just
above the ferromagnetic transition temperature~\cite{rivas2006}.
Moreover, Rairigh {\it et al.}~\cite{rairigh2007} observed a
``colossal magnetocapacitance'' coinciding with a regime of phase
separation between magnetic metal and charge-ordered insulator.
Magnetic materials have, of course, an intrinsic magneto-capacitive
coupling, but we suggest that the magnetodielectric response is
strongly enhanced in such a phase-separated composite, where the
magnetized domains combine to produce a large internal magnetic
field which acts on the transport currents.

To conclude, we have shown that a unique magnetic-field-dependent
dielectric resonance is produced by a strongly inhomogeneous media
and, as such, may be used as a probe of inhomogeneities or as a
magnetic sensor.

\begin{acknowledgements}
We are grateful to Neil Mathur and Gustau Catalan for fruitful
discussions. This research was supported in part by the National
Science Foundation under Grant Number DMR-0645461. 
\end{acknowledgements}


\end{document}